\documentclass[a4paper,10pt]{article}
\pdfoutput=1 

\usepackage{jcappub} 
\usepackage{amssymb}
\usepackage{amsmath,amssymb,graphicx}
\usepackage{graphicx}
\usepackage{dcolumn}
\usepackage{bm}
\def\be{\begin{equation}}
\def\ee{\end{equation}}
\def\bea{\begin{eqnarray}}
\def\eea{\end{eqnarray}}

\usepackage{slashed}
\usepackage{gensymb}
\def\lsim{\raise0.3ex\hbox{$\;<$\kern-0.75em\raise-1.1ex\hbox{$\sim\;$}}}
\def\gsim{\raise0.3ex\hbox{$\;>$\kern-0.75em\raise-1.1ex\hbox{$\sim\;$}}}

\newcommand{\mat}[1]{\begin{pmatrix} #1 \end{pmatrix}}
\newcommand{\nn}{\tilde{\nu}}

\usepackage{pstricks}

\usepackage{epsfig}

\title{\boldmath Dark Matter in $B-L$ Supersymmetric Standard Model with Inverse Seesaw}

\author[a,b]{W. Abdallah}
\author[a]{and S. Khalil}

\affiliation[a]{\small Center for Fundamental Physics, Zewail City of Science and Technology, 6 October City, Giza 12588, Egypt.}
\affiliation[b]{Department of Mathematics, Faculty of Science, Cairo University, Giza 12613, Egypt.}

\emailAdd{wabdallah@zewailcity.edu.eg}
\emailAdd{s.khalil@zewailcity.edu.eg}
\abstract{We show that the $B-L$ Supersymmetric Standard Model with Inverse Seesaw (BLSSMIS) provides new Dark Matter (DM) candidates (lightest right-handed sneutrino and lightest $B-L$ neutralino) with mass of order few hundreds GeV, while most of other SUSY spectrum can be quite heavy, consistently with the current Large Hadron Collider (LHC) constraints. 
We emphasize that the thermal relic abundance and direct detection experiments via relic neutralino scattering with nuclei impose stringent constraints on the $B-L$ neutralinos.  
These constraints can be satisfied by few points in the parameter space where the $B-L$ lightest neutralino is higgsino-like, which cannot explain the observed Galactic Center (GC) gamma-ray excess measured by Fermi-LAT.
The lightest right-handed sneutrino  DM is analysed. We show that for a wide region of parameter space the lightest right-handed sneutrino, with mass between $80$~GeV and $1.2$~TeV, can satisfy the limits of the relic abundance and the scattering cross with nuclei. We also show that the lightest right-handed sneutrino with mass ${\cal O}(100)$~GeV can account for the observed GC gamma-ray results.}
\begin{document}
\maketitle
\flushbottom
\section{Introduction}
Non-vanishing neutrino masses and the existence of non-baryonic DM are firm evidences of a new physics beyond not only the Standard
Model (SM) but also the Minimal Supersymmetric Standard Model (MSSM). The neutrinos are massless in the MSSM and the combined constraints from the LHC and the DM search limits rule out most of its parameter space, so it becomes very challenging for the MSSM Lightest Supersymmetric Particle (LSP) to remain a viable candidate of the DM \cite{Abdallah:2015hza,Chakraborti:2014fha,Roszkowski:2014wqa,Calibbi:2014lga}. 
Because of this, the $B-L$ (Baryon minus Lepton number) extension of the MSSM (BLSSM) is a well motivated scenario of new physics beyond the SM.  

In the BLSSM, (heavy) right-handed neutrino superfields are introduced in order to implement seesaw mechanism, which provides an elegant solution
for the existence and smallness of the (light) left-handed neutrino masses. TeV scale BLSSM with Type-I seesaw requires a finely tuned small neutrino Yukawa coupling ($ \lsim 10^{-7}$) \cite{Khalil:2006yi,Basso:2008iv,Basso:2009gg,Basso:2010yz,Basso:2010as,Majee:2010ar,Li:2010rb,Perez:2009mu,Emam:2007dy,PublicPapers}. 
However, if one implements Inverse Seesaw (IS) mechanism, neutrino Yukawa coupling does not have to be small and it can even be comparable to the top quark 
Yukawa coupling \cite{Khalil:2010iu}. One of the interesting feature of the BLSSMIS is that the SM-like Higgs boson mass gets a significant one loop radiative
corrections due to the right-handed (s)neutrinos~\cite{BLMSSM-Higgs},  which alleviate the so-called `little hierarchy problem' of the minimal SUSY realization, whereby the currently measured mass of the SM-like Higgs mass is very near
its absolute upper limit predicted theoretically, of $130$~GeV~\cite{Aad:2012tfa}. Furthermore, it was pointed out that in the BLSSMIS the scale of $B-L$ symmetry breaking can be naturally related to 
the SUSY breaking scale for a wide region of parameter space \cite{Khalil:2016lgy}.

The BLSSMIS provides new DM candidates that may account for the relic density with no conflict
with other experimental/phenomenological constraints. DM in extended MSSM models recently drew a lot of attention \cite{Guo:2013sna,DeRomeri:2012qd,Basso:2012gz,Khalil:2011tb,Khalil:2008ps,Cerdeno:2004xw}. In this paper, we consider the scenario where the extra
$B-L$ neutralinos (three extra neutral fermions: $U(1)_{B-L}$
gaugino and two extra higgsinos) or the lightest right-handed sneutrino can be the LSP, hence a stable and viable candidate for cold
DM. We examine the thermal relic abundance of these particles and discuss their direct and indirect detections if they form part of our galactic halo.
Our analysis is a completion for the work done in refs.~\cite{Khalil:2011tb,Khalil:2008ps}

The paper is organized as follows. In section~2 we define our BLSSMIS
model and present the possibility that the LSP could be the lightest $B-L$ neutralino or lightest right-handed sneutrino.
Section~3 is devoted to the study of $B-L$ neutralino DM, in particular thermal relic abundance, direct detection, and GC gamma-rays excess are investigated. 
In section~4 we analyse the right-handed sneutrino DM and emphasize that, unlike the scenario of lightest neutralino, there are many points of parameter space with lightest sneutrino mass between $80$~GeV and $1.2$~TeV are allowed. Also a significant number of these points predict spin-independent scattering cross sections that can be probed in near future. Finally, we show that lightest right-handed sneutrino with mass $\sim100$~GeV can account for GC gamma-ray excess. Our conclusions are given in section~5.
\section{Lightest Supersymmetric Particle in the BLSSMIS}
TeV scale BLSSMIS is based on the gauge group $SU(3)_C\times SU(2)_L\times U(1)_Y\times
U(1)_{B-L}$, where the $U(1)_{B-L}$ is spontaneously broken by 
chiral singlet superfields $\hat{\eta}_{1,2}$ with $B-L$ charge $=\pm 1$.
As in conventional $B-L$
model, a gauge boson $Z'$ and three chiral singlet
superfields $\hat{\nu}^c_i$ with $B-L$ charge $=-1$ are introduced for
the consistency of the model. Finally, three chiral singlet
superfields $S_1$ with $B-L$ charge $=+2$ and three chiral singlet
superfields $S_2$ with $B-L$ charge $=-2$ are considered to
implement the inverse seesaw
mechanism \cite{Khalil:2010iu}. The superpotential of this model is given by%
\be
W =  Y_u\hat{Q}\hat{H}_2\hat{U}^c + Y_d \hat{Q}\hat{H}_1\hat{D}^c+ Y_e\hat{L}\hat{H}_1\hat{E}^c+Y_\nu\hat{L}\hat{H}_2\hat{\nu}^c+Y_S\hat{\nu}^c\hat{\eta}_1\hat{S}_2 +\mu\hat{H}_1\hat{H}_2+ \mu'\hat{\eta}_1\hat{\eta}_2. 
\label{superpotential}
\ee
Since the chiral singlet superfields $\hat{\eta}_2$ and $\hat{\nu}^c$ have the same $B-L$ charge,  a discrete symmetry should be imposed to distinguish them and to prohibit other terms beyond those given in eq.~(\ref{superpotential}).  Also in this scenario, the light neutrino masses are assumed to be related to a
small mass term $\mu_s S^2_2$, with $\mu_s\sim {\cal O}(1)$ {\rm
KeV}. This mass term can be generated through a
non-renormalisable higher order interaction $\lambda^3\eta_1^4 S^2_2/M^3$, where $M$ is the mass of a heavy state whose loops
generate this term and it could be of order ${\cal O}(10^3)$~GeV if the coupling $\lambda$ associated to this interaction is $\sim {\cal O}(0.1)$. 
The relevant soft SUSY breaking terms, assuming the usual
universality assumptions at the Grand Unification Theory (GUT) scale, are given by
\bea - {\cal L}_{\rm soft} &=&
m_0^{2}\sum_{\phi} \vert \phi \vert^2 +Y^A_u\tilde{Q}\hat{H}_2\tilde{U}^c + Y^A_d \tilde{Q}H_1\tilde{D}^c+ Y^A_e\tilde{L}H_1\tilde{E}^c+
Y_{\nu}^{A}\tilde{L}H_{2}\tilde{\nu}^{c}+Y_{S}^{A}\tilde{\nu}^{c}\eta_1 \tilde
{S}_2 \nonumber\\ 
&+&\left[B \left(\mu H_1 H_2 + \mu' \eta_1 \eta_2\right) 
+\frac{1}{2}m_{1/2}\left(\tilde{g}^a \tilde{g}^a+\tilde{W}^a \tilde{W}^a+ \tilde{B}\tilde{B} + \tilde{B'}\tilde{B'}\right)+ h.c.\right],
\label{soft}
\eea%
where the sum in the first term runs over $\phi=\tilde{Q},\tilde{U},\tilde{D},\tilde{L},\tilde{E},\tilde{\nu},\tilde{S}_{1,2},H_{1,2},\eta_{1,2}$
and $(Y_f^A)_{ij}\equiv (Y_f)_{ij} A_{ij}$ ($f=u,d,e,\nu,S$) is the trilinear scalar interaction coupling associated with fermion
Yukawa coupling. The $B-L$ symmetry can be radiatively broken by
the non-vanishing Vacuum Expectation Values (VEVs): $\langle
\eta_1 \rangle = v'_1$ and $\langle \eta_2 \rangle = v'_2$. We define $\tan{\beta'}$ as the ratio of these VEVs ($\tan{\beta'}=v'_1/v'_2$) in analogy to the MSSM VEVs ($\tan{\beta}=v_2/v_1$) \cite{Khalil:2016lgy,Khalil:2007dr}.

It worth noting that the most general kinetic Lagrangian of the BLSSM allows for $U(1)_Y$ and $U(1)_{B-L}$ gauge kinetic mixing. This mixing can be absorbed in the covariant derivative redefinition, where the gauge coupling matrix will be transformed as follows:
\be
G = \left(\begin{array}{cc}
  g_{_{YY}} & g_{_{YB}}\\
  g_{_{BY}} & g_{_{BB}}\\
\end{array}%
\right) ~~ \Longrightarrow ~~ \tilde{G} = \left(\begin{array}{cc}
  g_1& \tilde{g}\\
  0 & g_{_{BL}}\\
\end{array}%
\right) , %
\ee
where 
\be
g_1 = \frac{g_{_{YY}} g_{_{BB}} - g_{_{YB}} g_{_{BY}}}{\sqrt{g_{_{BB}}^2 + g_{_{BL}}^2}}=g_2,~~~~g_{_{BL}} =\sqrt{g_{_{BB}}^2 + g_{_{BL}}^2},~~~~\tilde{g} = \frac{g_{_{YB}} g_{_{BB}} + g_{_{BY}} g_{_{YY}}}{\sqrt{g_{_{BB}}^2 + g_{_{BL}}^2}},
\ee
where $g_1$ and $g_2$ are $U(1)_Y$ and $SU(2)_L$ gauge couplings, respectively, and they are equal at the GUT scale due to the gauge coupling unification condition.

In this basis, after the $B-L$ and electroweak symmetry breaking, one finds
\be
M_Z^2 = \frac{1}{4} (g_1^2 +g_2^2) v^2,  ~~~~ M_{Z'}^2 = g_{_{BL}}^2 v'^2 + \frac{1}{4} \tilde{g}^2 v^2,
\ee
where 
\begin{equation}
{v=\sqrt{v_1^2+v_2^2}\simeq 246 ~ \text{GeV~~and~~}v'=\sqrt{v'^2_1+v'^2_2}}.
\end{equation}
Furthermore, the mixing angle between $Z$ and $Z'$ is given by 
\be 
\tan 2 \theta' = \frac{2 \tilde{g}\sqrt{g_1^2+g_2^2}}{\tilde{g}^2 + 4 (\frac{v'}{v})^2 g_{_{BL}}^2 -g_2^2 -g_1^2}.
\ee
\subsection{Lightest Neutralino}
\begin{figure}[t!]
\begin{center}
\includegraphics[width=8cm,height=6cm]{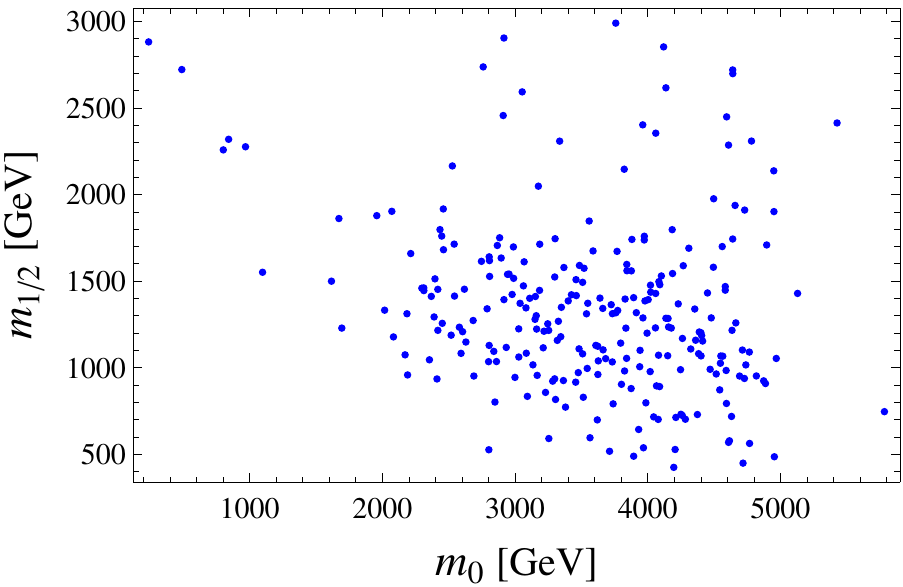}
\caption{The region in $m_0-m_{1/2}$ plane, where the lightest neutralino is $B-L$ higgsino-like with mass:  $m_{\tilde{\eta}_{2}} \simeq \mu' \lsim 600$~GeV. Here $A_0\in[-4,4]$~TeV and $\tan{\beta}\in[3,50]$ and all the LHC constraints are satisfied. } 
\label{m0m12}
\end{center}
\end{figure}

Now, we consider the neutralino sector in the BLSSMIS. In this model, the neutralinos $\tilde{\chi}^0_i $ ($i=1,\dots,7$) are
the physical (mass) superpositions of three fermionic partners of
neutral gauge bosons, are called gauginos $\tilde{B}$ (bino), 
$\tilde{W}^3$ (wino) and $\tilde{B'}$ ($B'$ino), in addition to the fermionic partners of neutral MSSM Higgs ($\tilde{H}_1^0$, and $\tilde{H}_2^0 $.) and the fermionic partners of $B-L$ scalar bosons ($\tilde{\eta}_1$, and $\tilde{\eta}_2 $). The $7\times 7$ neutralino mass matrix is given by
\be
{\cal M}_7({\tilde B},~{\tilde W}^3,~{\tilde
H}^0_1,~{\tilde H}^0_2,~{\tilde B'},~{\tilde \eta_1},~{\tilde
\eta_2}) \equiv \left(\begin{array}{cc}
{\cal M}_4 & {\cal O}\\
 {\cal O}^T &  {\cal M}_3\\
\end{array}\right),
\ee%
where the ${\cal M}_4$ is the MSSM neutralino mass matrix \cite{Haber:1984rc,Gunion:1984yn,ElKheishen:1992yv,Guchait:1991ia}, while
${\cal M}_{3}$ is $3\times 3$ additional $B-L$ neutralino mass matrix and the off-diagonal $3\times 3$ matrix ${\cal O}$ are given by%
\be%
{\cal M}_3 = \left(\begin{array}{ccc}
M_{B'} & -g_{_{BL}}v'_1  & g_{_{BL}}v'_2 \\
-g_{_{BL}}v'_1 & 0 & -\mu'  \\
g_{_{BL}}v'_2 & -\mu' & 0\\
\end{array}\right),~~~~~~{\cal O} = \left(\begin{array}{ccc}
\frac{1}{2}M_{BB'} &~~~0~~~& 0 \\
0 & 0 & 0  \\
-\frac{1}{2}\tilde{g}v_1 &~~~0~~~& 0\\
\frac{1}{2}\tilde{g}v_2&~~~0~~~&0\\
\end{array}\right),\label{mass-matrix.1} 
\ee
where $M_{B'}$ is $B'$ino mass (equals $m_{1/2}$ at the GUT scale) and $M_{BB'}$ is the mass mixing term of $\tilde{B}$ and $\tilde{B'}$ (equals zero at the GUT scale).    
Note that the elements of the matrix $\cal O$ vanish identically if $\tilde{g}=0$. In this case, one diagonalises the real matrix ${\cal M}_{7}$ with a symmetric mixing matrix $V$ such as
\be V{\cal
M}_7V^{T}={\rm diag}(m_{\tilde\chi^0_i}),~~i=1,\dots,7.\label{general} \ee In
these conditions, the LSP has the following decomposition 
\be 
\tilde\chi^0_1=V_{11}{\tilde B}+V_{12}{\tilde
W}^3+V_{13}{\tilde H}^0_1+V_{14}{\tilde
H}^0_2+V_{15}{\tilde B'}+V_{16}{\tilde \eta_1}+V_{17}{\tilde
\eta_2}. 
\ee 
The LSP is called pure $B'$ino ($\tilde{B'}$) if $V_{15}\sim1$ and $V_{1i}\sim0$ for $i\neq5$, and pure $B-L$ higgsino $\tilde\eta_{1(2)}$ if $V_{16(7)}\sim1$ and all the other coefficients are close to zero. However, as can be seen from eq.~(\ref{mass-matrix.1}), the off-diagonal elements $({\cal M}_3)_{12,13}$ and  $({\cal M}_3)_{21,31}$ are not suppressed. Therefore, unless $\mu'$ is very large, the lightest $B-L$ neutralino is a mixed between $B'$ino and $\tilde{\eta}_{1,2}$, i.e., $V_{15}, V_{16}$ and $V_{17}$ are not negligible and even comparable. With $\tan\beta'\simeq 1$, i.e.,  $v'_1\simeq v'_2 \simeq v'/\sqrt{2}$, one finds the following eigenvalues of $B-L$ neutralinos mass matrix ${\cal M}_3$:
\begin{eqnarray}
m_{\tilde{\chi}^0_5}&\simeq& \mu', \\
m_{\tilde{\chi}^0_{6,7}}&\simeq& \frac{1}{2}\left(M_{B'}+\mu'\mp \sqrt{(M_{B'}-\mu')^2+4 g_{_{BL}}^2 v'^2} \right).
\end{eqnarray}
Therefore, if $\mu' \gg M_{B'}$, one obtains $m_{\tilde{\chi}^0_5} \simeq  \mu'$, $m_{\tilde{\chi}^0_6} \simeq  \frac{1}{2}(M_{B'} + 2 \mu') \sim \mu'$, and $m_{\tilde{\chi}^0_7} \simeq  \frac{1}{2} M_{B'}+ g_{_{BL}}^2 v'^2/\mu'$, so in this case the LSP could be $B'$ino with mass of order $M_{B'}$. In addition, if $\mu' \ll M_{B'}$, the LSP would be $\tilde{\chi}^0_5$ (mainly $\tilde{\eta}_2$) with mass of order $\mu'$ and no degeneracy with $\tilde{\eta}_1$ (unlike the case of higgsino-like LSP in the MSSM).  
Recall that the $\mu'$ parameter is determined by the $B-L$ minimization condition (for $\tilde{g}=0)$ as follows:
\be
\mu'^2 = \frac{m^2_{\eta_2} - m^2_{\eta_1} \tan^2 \beta'}{\tan^2\beta' -1} -\frac{1}{4} M_{Z'}^2.
\label{mu-prime}
\ee
Thus, the typical value of $\mu'$ is of order $v' \sim {\cal O}(1)$~TeV. However, as in case of $\mu$-parameter in the MSSM, it is possible to find $\mu' \sim {\cal O}(100)$~GeV. In fact,  this possibility is even larger in the BLSSM since $\tan \beta'$ is close to one and hence a significant cancellation among the terms in the right-hand-side of eq.~(\ref{mu-prime}) may occur. In figure \ref{m0m12}, we show the region of $m_0-m_{1/2}$ plane that leads to $\tilde{\eta}_{2}$ is the LSP with $\vert V_{17} \vert^2 > 0.7$ and $m_{\tilde{\eta}_{2}} \simeq \mu' \lsim 600$~GeV. 

From this figure, it is remarkable that most of the points that lead to $m_{\tilde{\eta}_2} \lsim 600$~GeV correspond to heavy $m_0$ and  $m_{1/2}$, i.e., the usual MSSM spectrum would be quite heavy and can not be probed at the LHC in run I \cite{SUSYRUN1} and may be run II as well.   
\subsection{Lightest Right-handed Sneutrino}
Now we turn to the sneutrino mass matrix. If we write $\tilde{\nu}_{L}$, $\tilde{\nu}_{R}$ and  $\tilde{S}_2$ as 
\be
\tilde{\nu}_L  =  \frac{1}{\sqrt{2}} \left(\tilde{\nu}_L^+ + i\; \tilde{\nu}_L^- \right),~~~~\tilde{\nu}_R  =  \frac{1}{\sqrt{2}} \left( \tilde{\nu}_R^+ + i\; \tilde{\nu}_R^- \right),~~~~\tilde{S}_2 =  \frac{1}{\sqrt{2}} \left( \tilde{S}_2^+ + i\; \tilde{S}_2^- \right),
\ee
then the sneutrino mass matrix can be written as 
\begin{eqnarray}
M_{\nn}^2 = \mat{{\cal M}^2_+ & 0\\0 &{ \cal M}^2_-},
\end{eqnarray}
where the CP-even/odd sneutrino mass matrix (for $\tilde{g}=0$) is given by 

{\small\fontsize{8}{8}\selectfont{
\be
{\cal M}^2_{\pm}=\mat{
m_{\tilde{L}}^2+m_D^2+\frac{1}{2}(M_Z^2\cos 2\beta+M_{Z'}^2\cos 2\beta') & \pm m_D(A_{\nu}+\mu\cot\beta) & m_D M_R\\\\
\pm m_D(A_{\nu}+\mu\cot\beta) & m_{\tilde{\nu}_R}^2+m_D^2+M_R^2-\frac{1}{2}M_{Z'}^2\cos 2\beta' & \pm M_R (A_S+\mu'\cot\beta')\\\\
m_D M_R & \pm M_R (A_S+\mu'\cot\beta') & m_{\tilde{S}}^2+M_R^2+M_{Z'}^2\cos 2\beta'
}.\nonumber
\label{Mpm}
\ee
}}
\begin{figure}[t]
\begin{center}
\includegraphics[width=8cm,height=5.7cm]{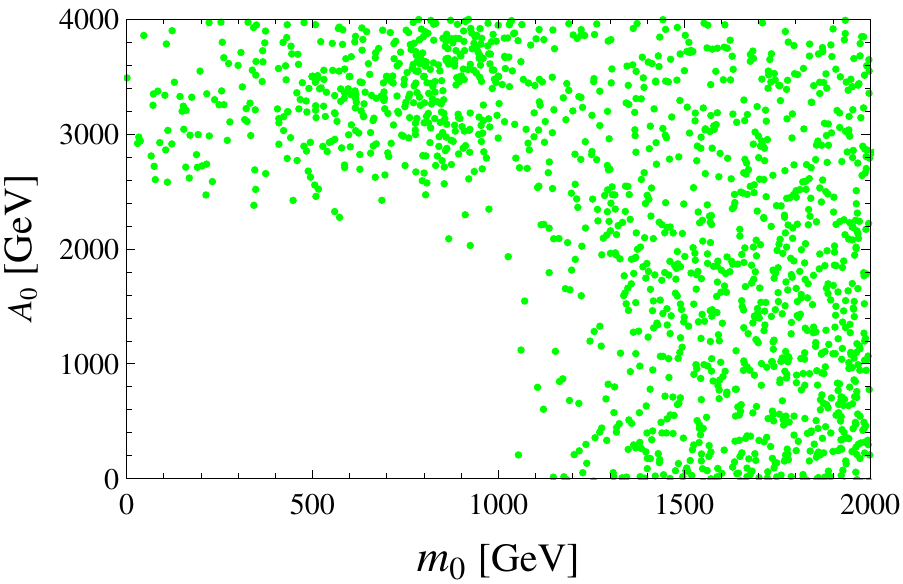}
\caption{The region of $m_0 - A_0$ plane that leads to $m_{\tilde{\nu}_1} \lsim 600$~GeV. Here, $m_{1/2}\in[0,3]$~TeV, $\tan{\beta}\in[3,50]$ and all the LHC constraints are satisfied. } 
\label{m0A0Msvsmaller600}
\end{center}
\end{figure}
The diagonalization of these matrices (especially with non-vanishing $\tilde{g}$) is not an easy task and can only be performed numerically. It turns out that the mass of the lightest CP-odd sneutrino, $\tilde{\nu}^-_{i}$, is almost equal to the mass of the lightest CP-even sneutrino, $\tilde{\nu}^+_{i}$. Also, both $\tilde{\nu}^{-}_{2,3}$ and $\tilde{\nu}^{+}_{2,3}$ are generated from the mixing between $\tilde{\nu}_R$ and $\tilde{S}_2$. In fact, since the off-diagonal elements $( {\cal M}^2_{\pm})_{12,13}$ and $( {\cal M}^2_{\pm})_{21,31}$ are much smaller than other elements, the matrices ${\cal M}^2_{\pm}$ can be approximately decomposed into $1\times 1$ (corresponds to left-handed sneutrinos) and $2\times 2$ (corresponds to right-handed sneutrinos) block diagonal matrices. Thus, for $\mu'$ and/or $A_S$ are of order $m_{\tilde{\nu}_R}$ and $M_R$, i.e., $\sim {\cal O}(1)$~TeV, one of the eigenvalues of $ {\cal M}^2_{\pm}$ is expected to be light. This lightest sneutrino can be of order ${\cal O}(100)$~GeV, as shown in figure \ref{m0A0Msvsmaller600}, where we display the region of $A_0 (\sim A_S) -m_0(\sim\mu'/2)$ plane that leads to lightest right-handed sneutrino with mass less than $600$~GeV.   
 
This figure, which is based on numerical calculation of the full sneutrinos mass matrix with non-vanishing $\tilde{g}$, confirms the above conclusion that $A_0$ and/or $m_0$ must be quite larger ($\sim$~TeV), in order to obtain a light right-handed sneutrino ($\sim {\cal O} (100)$~GeV).  This type of LSP could be an interesting example of DM candidate with quite heavy SUSY spectrum, required to satisfy current/future LHC constraints. 
\begin{figure}[t]
\begin{center}
\includegraphics[width=8cm,height=6cm]{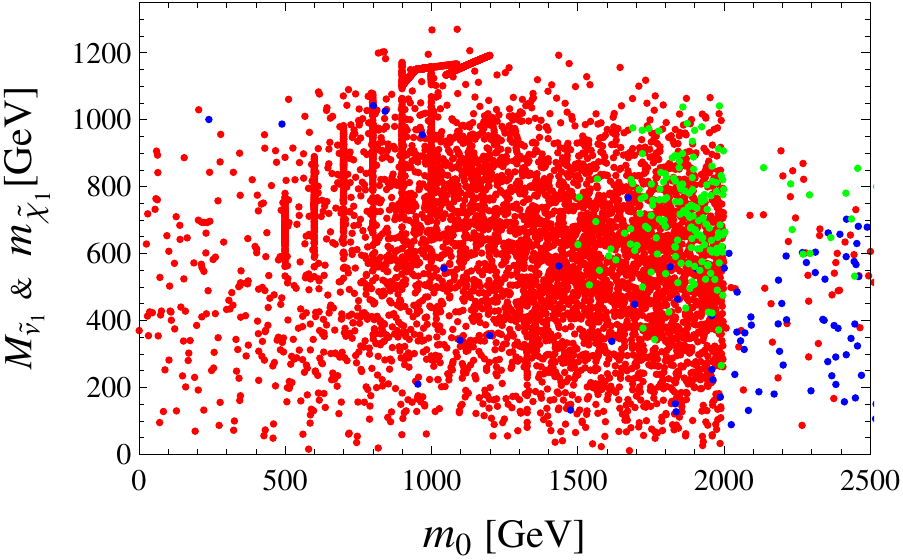}
\caption{The right-handed sneutrino/$B-L$ neutralino LSP mass versus the scalar soft mass $m_0$ for $m_{1/2}\in[0,3]$~TeV, $A_0\in[-4,4]$~TeV and $\tan{\beta}\in[3,50]$ after imposing the Higgs and gluino mass constraints. Red points indicate to the right-handed sneutrino LSP, while the green and blue points correspond to $\tilde{B'}$ and $\tilde{\eta}_2$ LSP, respectively.  } 
\label{m0mchimsnu}
\end{center}
\end{figure}

The lightest sneutrino $\tilde{\nu}_1$ (either it is CP-even sneutrino, $\tilde{\nu}^{\rm R}_1$, or CP-odd sneutrino, $\tilde{\nu}^{\rm I}_1$) can be expressed in terms of $\tilde{\nu}^+_L$, $\tilde{\nu}^+_R$, $\tilde{S}^+_2$ (in case of it is CP-even) as 
\be 
\tilde{\nu}_1 = \sum_{i=1}^3 R_{1i} (\tilde{\nu}^+_L)_i +  \sum_{j=1}^3 R_{1j} (\tilde{\nu}^+_R)_j +  \sum_{k=1}^3 R_{1k} (\tilde{S}^+_2)_k.
\label{Gamma}
\ee
If one defines the decomposition coefficients as $R_{1l} \equiv \{R_{1i}, R_{1j},R_{1k}\}$, i.e.,  $l=1,\dots, 9$, then it is easily to show that the typical value of $R_{1l}$, with diagonal $A_S$ and $M_R$, is given by $R_{1l} =\frac{1}{\sqrt{2}} \{ 0,0,0,1,0,0,1,0,0\}$. This confirms that the lightest sneutrino is mainly right-handed (i.e., a combination of $\tilde{\nu}^+_R$ and $\tilde{S}^+_2$). 
\subsection{Lightest $B-L$ Neutralino Versus Lightest Right-handed Sneutrino}
As intimated, the BLSSMIS has more candidates for DM, in addition to the usual lightest MSSM neutralino (bino, wino, higgsino)-like. In particular, the lightest $B-L$ neutralino ($\tilde{B'}$, $\tilde{\eta}_{2}$)-like and lightest right-handed sneutrino can be strong candidates. Here, we investigate the possibility that the $B-L$ lightest neutralino or lightest right-handed sneutrino makes up the LSP and hence becomes the main candidate for the DM.  In figure \ref{m0mchimsnu}, this conclusion is emphasized by performing a scan over the relevant parameter space of the BLSSMIS, namely $m_0\in[0,5]$~TeV, $m_{1/2}\in[0,3]$~TeV, $A_0\in[-4,4]$~TeV and $\tan{\beta}\in[3,50]$ with imposing the Higgs and gluino mass constraints \cite{gluino search}.

From this figure, one can easily notice that the lightest right-handed sneutrino (red points) is most likely to be the LSP for a wide region of parameter space. The $B'$ino-like (green points) is the second possibility for the LSP, which occurs as explained above if $\mu' (m_0)$ is quite heavy. This explains the existence of the green points in the lower right corner of the plot, where $m_{1/2} < 1 $~TeV and $m_0 > 1$~TeV. Finally, the $B-L$ higgsino ($\tilde{\eta}_2$)-like (blue points) can be the LSP for a narrow region of parameter space, namely when $\mu'$ is lighter than $M_{B'}$.   
\section{$B-L$ Neutralino Dark Matter}
As advocated above, in the BLSSMIS the LSP is most likely pure right-handed sneutrino with a small chance of $B-L$ neutralino ($\tilde{B'}$ or $\tilde{\eta}_2$). In this section, we focus on the possibility of having $B-L$ neutralino as DM candidate and analyse its relic abundance, the constraints imposed on this region of parameter space for direct detection experiments and the Galactic Center Excess (GCE). 
\subsection{$B-L$ Neutralino Relic Abundance}
In studying the relic abundance, we consider the scenario of standard cosmological, where the LSP is assumed to be in thermal equilibrium with the SM particles in the early universe and decoupled when it was non-relativistic. Therefore, the density of the lightest $B-L$ neutralino (hereafter, we refer to it as $\tilde{\chi}_1$) can be obtained by solving the Boltzmann equation \cite{Kolb,Jungman:1995df}:\\ \vspace{-0.2cm}
\be 
\frac{d n_{\tilde{\chi}_1}}{dt}+3Hn_{\tilde{\chi}_1}=-\langle\sigma^{\rm ann}_{\tilde{\chi}_1}v\rangle
\left[(n_{\tilde{\chi}_1})^2-(n^{\rm eq.}_{\tilde{\chi}_1})^2\right],
\label{boltzmann}
\ee %
where $n_{\tilde{\chi}_1}$ is the number density of $\tilde{\chi}_1$  with
$\rho_{\tilde{\chi}_1}=m_{\tilde{\chi}_1}n_{\tilde{\chi}_1}$. One
usually defines
$\Omega_{\tilde{\chi}_1}=\rho_{\tilde{\chi}_1}/\rho_{c}$, where
$\rho_{c}$ is the critical mass density. In addition,
$\langle\sigma^{\rm ann}_{\tilde{\chi}_1}v\rangle$ is the thermal averaged of the
total cross section for $\tilde{\chi}_1$ annihilation into SM
lighter particles times the DM relative velocity $v$.  The relic density of the DM $\tilde{\chi}_1$ is given by 
\begin{equation}
\Omega h^2_{\tilde{\chi}_1}=\frac{2.1 \times 10^{-27 }~{\rm cm}^3~{\rm s}^{-1}}{\langle\sigma^{\rm ann}_{\tilde{\chi}_1}v\rangle}\left(\frac{x_F}{20}\right)\left(\frac{100}{g_*(T_F)}\right)^\frac{1}{2},
\end{equation}
where $g_* \simeq {\cal O}(100)$ is the degrees of freedom and $x_F =m_{\tilde{\chi}_1}/T_F\simeq {\cal O}(20)$ at the freeze out temperature, $T_F$.  

The relevant interactions of the $B-L$ neutralinos-like LSP, $\sum_{i=5}^7 |V_{1i}|^2 \sim 1~ {\rm and }~ V_{1j}\sim 0~{\rm for}~j =1,\dots,4$, are given by
\bea
{\cal L}_{\tilde{\chi}_1}
 &\simeq&\!\! 2i~g_{_{BL}}\overline{\tilde{\chi}_1} h' \left[
V_{15}^*\left(\Gamma_{33}V_{16}^*-\Gamma_{34}V_{17}^*\right) P_L+V_{15}\left(\Gamma_{33}V_{16}-\Gamma_{34}V_{17}\right) P_R\right]\tilde{\chi}_1\nonumber\\[0.3cm]
&+&\!\! 2~g_{_{BL}}\overline{\tilde{\chi}_1} A'\left[
V_{15}^*\left(\Lambda_{33}V_{16}^*-\Lambda_{34}V_{17}^*\right)P_L-V_{15}\left(\Lambda_{33}V_{16}-\Lambda_{34}V_{17}\right)P_R\right] \tilde{\chi}_1\nonumber\\[0.3cm]
&-&\!\! i~g_{_{BL}}\left(|V_{16}|^2-|V_{17}|^2\right)\overline{\tilde{\chi}_1}\slashed{Z'}\gamma_5\tilde{\chi}_1.
\eea
\begin{figure}[t]
\begin{center}
\includegraphics[width=6cm,height=3.5cm]{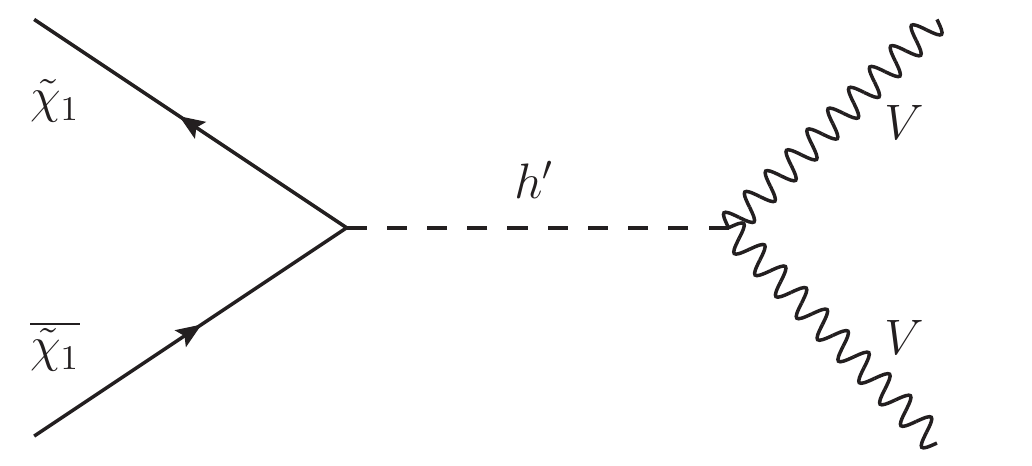}~~~~~~\includegraphics[width=6cm,height=3.5cm]{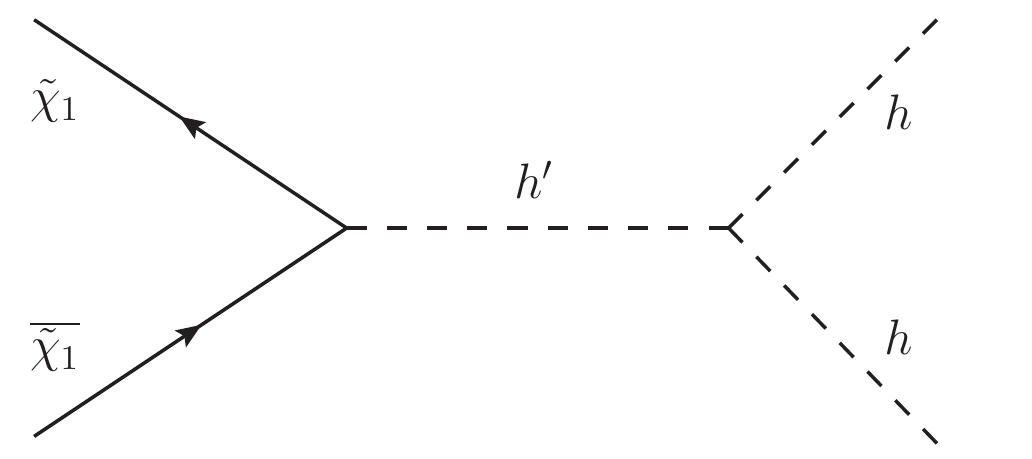}
\caption{Feynman diagrams of the dominant annihilation channels of the $B-L$ lightest neutralino $\tilde{\chi}_1$ into the SM vector bosons $(V=W,Z)$ and the SM-like Higgs $h$ mediated by the lightest $B-L$ CP-even Higgs.} \label{annihilationchi}
\end{center}
\end{figure}
In figure \ref{annihilationchi}, we show the Feynman diagrams of the dominant annihilation channels of the lightest $B-L$neutralino into $W W, ~ZZ, ~ h h $, through the lightest $B-L$ CP-even Higgs boson. Other annihilation channels mediated by $Z'/A'$ are suppressed due to the heavy mass of these particles.

\begin{figure}[t]
\begin{center}
\includegraphics[width=8cm,height=6cm]{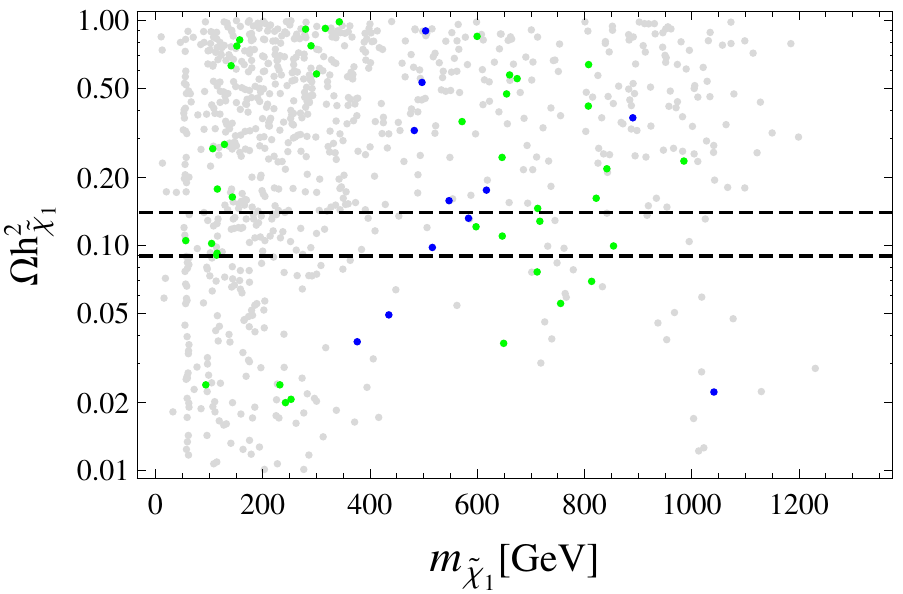}
\caption{The thermal relic abundance of $B-L$ neutralinos, $\tilde{B'}$-like (green points) and $\tilde{\eta}_2$-like (blue points), LSP as a function of their masses. Horizontal lines correspond to the Planck limits on DM abundance. The gray points indicate to the excluded points by the LHC and LEP constraints.} 
\label{MchiOGh2ALL}
\end{center}
\end{figure}

In figure \ref{MchiOGh2ALL}, we show the constraint from the $\Omega h^2_{\tilde{\chi}_1}$ observed limits as function of the lightest $B-L$ neutralino mass for $m_0,~m_{1/2} \in [100~{\rm GeV}, 3~{\rm TeV}]$, $A_0\in[-4,4]$~TeV, $\tan \beta \in[3, 50]$, and $\mu >0$. Here we adopt $2 \sigma$ results reported recently by Planck satellite \cite{Ade:2015xua}, namely we assume
\be
0.09<\Omega h^2<0.14
\label{plancklimit}
\ee
As usual, the LHC constraints, in particular, the SM-like Higgs and gluino mass constraints, are imposed. We used micrOMEGAs \cite{Belanger:2014vza} to compute the complete relic abundance of $\tilde{\chi}_1$: $B'$ino-like (blue points) or $B-L$ higgsino-like (green points). As can be seen from this figure, the narrow range of the relic abundance limits impose stringent constraints on this type of DM candidates. One finds only three benchmark points with $B'$ino DM are allowed and few points with $B-L$ higgsino ($\tilde{\eta}_2$) DM are allowed. Note that the masses of allowed $\tilde{\eta}_2$ are larger than $100$~GeV and less than TeV. It is remarkable that these allowed points are much larger than the corresponding ones in the MSSM, where no point with bino-like is allowed and much less points for higgsino-like at very large $\tan \beta$ are allowed \cite{Abdallah:2015hza}. 
\subsection{Direct Detection Constraints on the $B-L$ Neutralino }
\begin{figure}[t]
\begin{center}
\includegraphics[width=8cm,height=6cm]{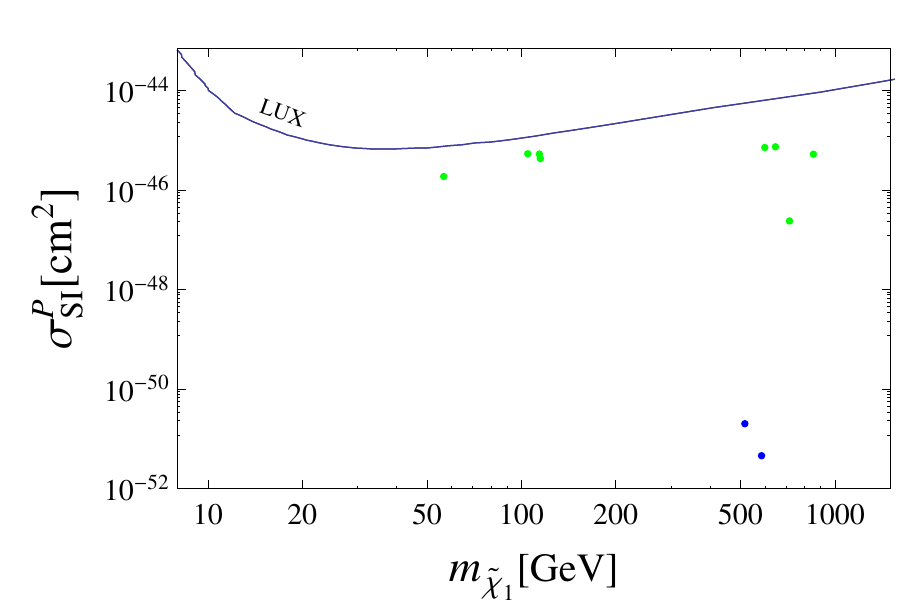}
\caption{Spin-independent scattering cross section of the $B-L$ lightest neutralino, $\tilde{B'}$-like (blue points) and $\tilde{\eta}_2$-like (green points), with a proton versus the mass of the LSP within the region allowed by all constraints (from the LHC and relic abundance). } 
\label{LUX}
\end{center}
\end{figure}
We now discuss the spin-independent DM scattering cross section of the allowed $B-L$ lightest neutralinos studied in the previous section. 
The effective scalar interactions of $B-L$ neutralino with up and down quarks are mainly given by $Z'$ exchange, {\it i.e.},  
\be%
{\cal L_{\text{scalar}}} = f_q \overline{\tilde{\chi}}_1 \tilde{\chi}_1 \, \bar{q} q, %
\label{scalar}
\ee%
where $f_q \propto g^2_{_{BL}}/M_{Z'}^2$, with $M_{Z'} > 2$~TeV. Therefore, the $\tilde{\chi}_1$ coupling to protons and neutrons $f_p$, $f_n$ (proportional to $f_u$ and $f_d$) 
are quite suppressed. The zero momentum transfer of scalar cross section of the neutralino scattering with nucleus is given by \cite{Jungman:1995df}%
\be%
\sigma^{\rm SI}_0 = \frac{4 m_r^2}{\pi} \left(Z f_p + (A-Z) f_n \right)^2,
\ee%
where $Z$ and $A-Z$ are the number of protons and neutrons, respectively, $m_r=m_N m_{\tilde{\chi}_1}/(m_N+m_{\tilde{\chi}_1})$, where $m_N$ is the nucleus mass. Thus, the differential scalar cross section for non-zero momentum transfer $q$ can be written as
\be%
\frac{d \sigma_{\rm SI}}{dq^2} = \frac{\sigma^{\rm SI}_0}{4 m_r^2 v^2}F^2(q^2),\; 0< q^2 < 4 m^2_r v^2,
\ee%
where $v$ is the velocity of the lightest neutralino and $F(q^2)$ is the form factor \cite{Jungman:1995df}. Therefore, the spin-independent scattering cross section of the LSP with a proton is given by
\be
\sigma_{\rm SI}^p=\int_0^{4 m^2_r v^2}\frac{d \sigma_{\rm SI}}{dq^2}\big{|}_{f_n=f_p} dq^2.
\ee
As intimated,  due to the suppression of the lightest $B-L$ neutralino scattering with proton, the spin-independent cross section of this scattering is expected to be very small. 
In figure \ref{LUX}, we display the spin-independent scattering cross section of the $B-L$ neutralinos LSP with a proton  after imposing the LHC and relic abundance constraints. It is clear that $\sigma_{\rm SI}^p$ results of $B-L$ neutralinos are less than the recent LUX bound \cite{Akerib:2015rjg}, blue curve, by at least two orders of magnitude. 
\subsection{$B-L$ Neutralino and Galactic Center $\gamma$-rays}
Searches for DM signal utilizing gamma-ray telescopes have attracted wide attention. The Large Area Telescope (LAT) on the Fermi Gamma-ray Space Telescope (FGST) mission has energy range from $20$~MeV to $300$~GeV. Therefore, it is very appropriate for searching for gamma-rays possibly produced by DM annihilation.  An excess of gamma-rays photons in 3--4~GeV range from the GCE is established by different Fermi-LAT observations \cite{Daylan:2014rsa,Ackermann:2015zua,Hooper:2013rwa}. These excess can be explained by DM particle with mass $\lsim {\cal O}(100)$~GeV and annihilation cross section of order $\langle \sigma^{\rm ann} v \rangle \simeq 10^{-26}~{\rm cm}^3~{\rm s}^{-1}$.

 The differential flux of gamma-rays over all annihilating channels of the DM particle (labeled by the subindex $i$) is described by 
 \be 
 \frac{d\Phi_\gamma(E_\gamma, \psi)}{dE_{\gamma}} = \sum_i \frac{dN^i_\gamma}{d E_\gamma} \frac{\langle \sigma^{\rm ann}_i v\rangle}{8\pi m_{\tilde{\chi}_1}^2}\frac{1}{\Delta\Omega}\int_{\Delta\Omega} d\Omega \int_{\rm los} \rho^2(r)~dl,\label{Eq:DFDM} 
 \ee
where $d N^i_\gamma/dE_\gamma$ is  the gamma-ray spectrum produced per annihilation $i$, which depends on the mass and the dominant annihilation channels of the DM particle and  is calculated by using PYTHIA \cite{Sjostrand:2006za}.
The astrophysical factor of eq.~(\ref{Eq:DFDM}) can be identified as:
\be
\langle J \rangle_{\Delta\Omega} = \int_{\Delta\Omega} d\Omega \int_{\rm los} \rho^2(r)~dl,
\ee
where the first integral is performed over the solid angle of the detector $\Delta\Omega=2\pi(1-\cos\psi)$ and the second integral is performed over the light-of-sight (los): from $0$ to the distance $l_{\rm max}$ between the sun and the edge of the halo, $l_{\rm max}(\psi)=r_{\odot}\cos{\psi}+\sqrt{r^2-r^2_{\odot}\sin^2{\psi}}$, where the radial distance $r$ from the GC is related to the distance $l$ from the sun to any point of the halo as follows: 
\be
r^2=\left(l^2+r^2_{\odot}-2 l r_{\odot}\cos{\psi}\right)^{\frac{1}{2}},
\ee
where $r_{\odot}=8.5$~kpc is the distance between the sun and the GC, $\psi$ is the angle observed relative to the direction of the GC and $\rho(r)$ is the DM density, which in case of a generalized Navarro-Frenk-White (NFW) halo profile with inner slope $\gamma$ is given by \cite{Navarro:1995iw,Navarro:1996gj} 
\be 
\rho(r) = \rho_0 \frac{(r/r_s)^{-\gamma}}{(1+r/r_s)^{3-\gamma}}, 
\ee
where $r_s = 20$ kpc is the scale radius, $\rho_0 =0.4~{\rm GeV}/{\rm cm}^3$ is the local DM density. Therefore, in the case of $\gamma=1.3$ and within a Region Of Interest (ROI) at galactic latitudes
 $2^{\circ} \leq |b| \leq 20^{\circ}$ and galactic longitudes $|l| \leq 20^{\circ}$, one finds the astrophysical factor $\langle J \rangle_{\Delta\Omega}$ of order ${\cal O}(10^{22})$~GeV$^2$~cm$^{-5}$ \cite{Daylan:2014rsa,Calore:2014xka}. The inner region $(|b|<2^{\circ})$ is avoided because this region of the sky that is most contaminated with strong gamma-ray point sources and with very large uncertainties in the diffuse emission.

The $B-L$ neutralinos (either $\tilde{B'}$ or $\tilde{\eta}_2$) cannot account for such gamma-ray excess for the following two reasons: $(i)$ The $B-L$ neutralino masses are constrained by relic abundance to be larger than $100$~GeV. $(ii)$ As the $B-L$ neutralinos are Majorana fermions, their annihilation cross section can be approximated as $\langle \sigma^{\rm ann} v \rangle \simeq b v^2$ (no s-wave), which should be of order $10^{-26}~ {\rm cm}^3~{\rm s}^{-1}$ at the freeze out temperature, where the velocity of the DM $v_F$ is of order $0.1~{\rm c}$, where $c$ is the speed of light. In our galactic halo the velocity of the DM particles is much smaller $(v \sim 10^{-3}~{\rm c})$. Therefore, the $B-L$ neutralino annihilation cross sections in the galactic halo are of order $10^{-30}~ {\rm cm}^3~{\rm s}^{-1}$, which are quite small to account for the GCE. 
\section{Right-handed Sneutrino Dark Matter}
We now turn to the lightest right-handed sneutrino LSP, $\tilde{\nu}_1$. As in the previous section, we will investigate the relic abundance constraints imposed on the parameter space in this scenario, then we analyse the prediction of this candidate for the direct detection and gamma-rays flux from the GC.  
\begin{figure}[t!]
\begin{center}
\vspace{-1.74cm}
\includegraphics[width=5cm,height=3cm]{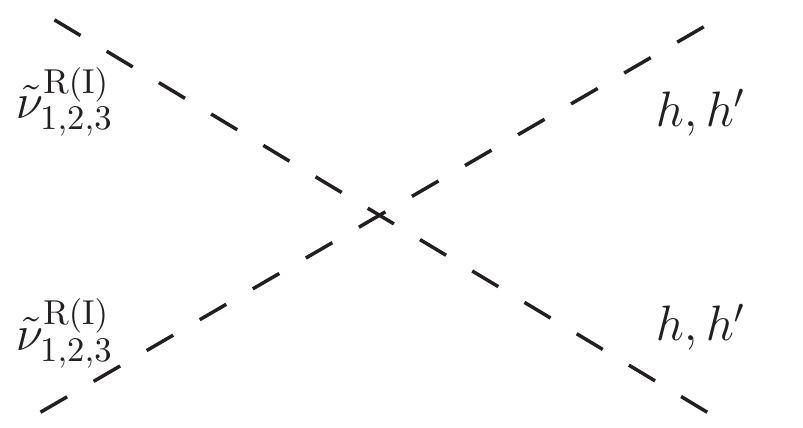} ~~~~~~\includegraphics[width=6cm,height=3cm]{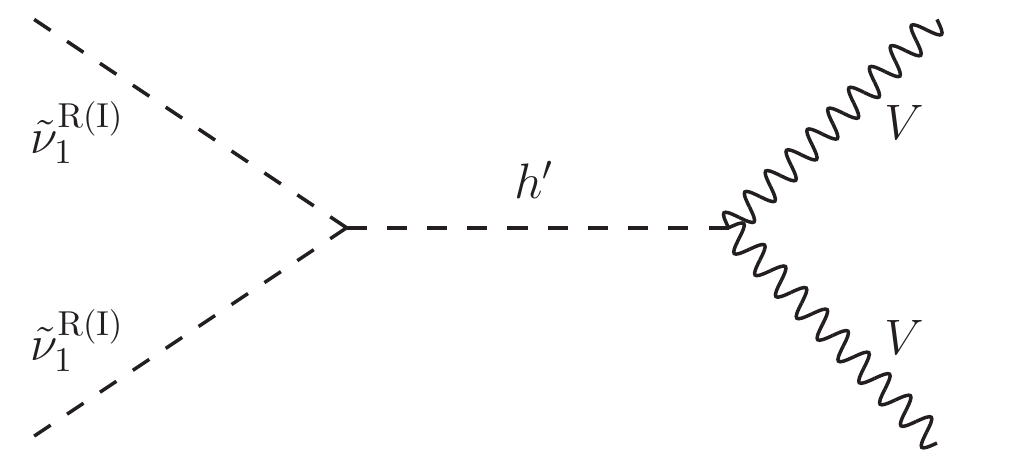}\\[0.3cm]
\includegraphics[width=5cm,height=3cm]{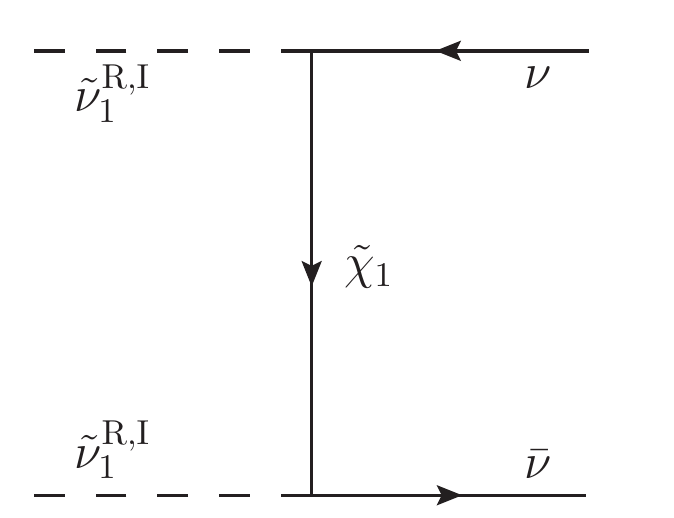} ~~~~~~ \includegraphics[width=6cm,height=3cm]{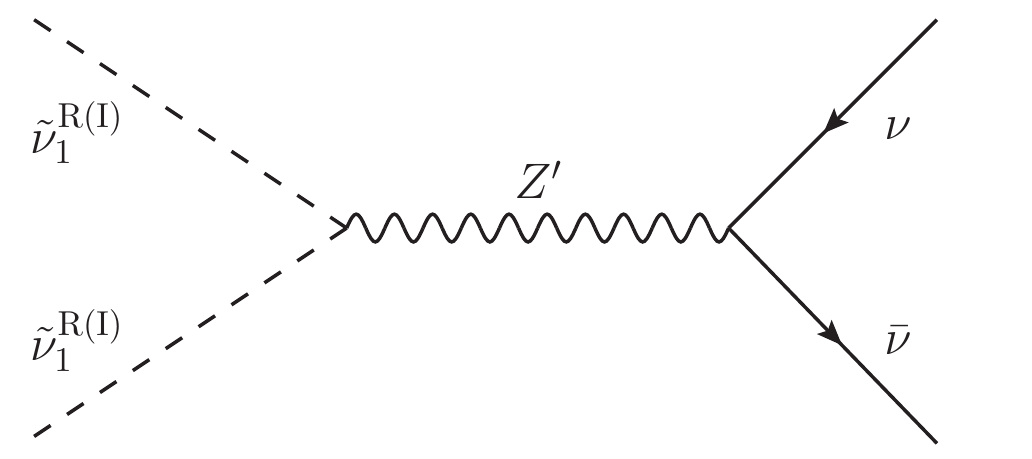}
\caption{Feynman diagrams of the dominant annihilation and co-annihilation channels of the right-handed sneutrino $\tilde{\nu}_1$ into the CP-even Higgs bosons $(h,~h')$, the SM vector bosons $(V=W,Z)$ and $\nu\bar\nu$.} \label{annihilationsnu}
\end{center}
\end{figure}
\subsection{Right-handed Sneutrino Relic Abundance}
The relevant interactions for dominant channels of right-handed sneutrino annihilation are given by: 
{\small\fontsize{8}{8}\selectfont{
\bea
{\cal L}_{\tilde{\nu}_1}\!\!\! &=&
i \!\!\!\!\sum_{n,m,l=1}^3 (\tilde\nu^{\rm R}_1)^2 |h|^2 \left[
\frac{\tilde{g}g_{_{BL}}}{4}  \left(\Gamma^2_{11}- \Gamma^2_{12}\right)-  \Gamma^2_{12} R^*_{1,l+3} R^*_{1,m+3} {Y_\nu}_{mn}{Y_\nu}_{ln}\right]\!\!+\! (\tilde\nu^{\rm R}_1)^2 |h'|^2 \left[\frac{ g^2_{BL}}{2}\left( \Gamma^2_{33}-\Gamma^2_{34}\right)- \Gamma^2_{33} {Y_S}^2_{n} \right]\nonumber\\
&+& i \sum_{n=1}^3 (\tilde\nu^{\rm R}_1)^2 h' \left[
\frac{1}{2} g_{_{BL}}^2 \left(v'_1 \Gamma_{33}- v'_2 \Gamma_{34}\right)-\sqrt{2} \mu' \Gamma_{34} R^*_{1,n+6} R^*_{1,n+3} {Y_S}_{n}
-\Gamma_{33} \left(
\sqrt{2} R^*_{1,n+6} R^*_{1,n+3} {T_S}_{n}+v'_1  {Y_S}^2_{n}
\right)\right]\nonumber\\
&+& i \sum_{n,m,l=1}^3 (\tilde\nu^{\rm R}_1)^2 h\left[\frac{1}{4} \tilde{g} g_{_{BL}}\left(v_1\Gamma_{11}-v_2\Gamma_{12}\right)-v_2 \Gamma_{12} R^*_{1,l+3} R^*_{1,m+3} {Y_\nu}_{mn} {Y_\nu}_{ln}\right]+\left[ (\tilde\nu^{\rm R}_1)^2 \to  (\tilde\nu^{\rm I}_1)^2~\text{and}~R_{ij} \to I_{ij}\right]\nonumber\\
&-& i \overline{\tilde{\chi}_1}~\tilde{\nu}^{\rm R}_1\nu^i_L\sum_{n=1}^3{\left[\frac{V^*_{15}}{2}g_{_{BL}}\left(R^*_{1,3+n}U^*_{i,3+n}+R^*_{1,6+n}U^*_{i,6+n}\right)+\frac{V^*_{16}}{\sqrt{2}}{Y_S}_{n}\left(R^*_{1,6+n}U^*_{i,3+n}+R^*_{1,3+n}U^*_{i,6+n}\right)\right]}\nonumber\\
&+& \overline{\tilde{\chi}_1}~\tilde{\nu}^{\rm I}_1\nu^i_L\sum_{n=1}^3{\left[\frac{V^*_{15}}{2}g_{_{BL}}\left(I^*_{1,3+n}U^*_{i,3+n}-I^*_{1,6+n}U^*_{i,6+n}\right)+\frac{V^*_{16}}{\sqrt{2}}{Y_S}_{n}\left(I^*_{1,6+n}U^*_{i,3+n}-I^*_{1,3+n}U^*_{i,6+n}\right)\right]}\nonumber\\
&+&\frac{1}{2} g_{_{BL}} \sum_{n=1}^3\tilde\nu^{\rm R}_1\tilde\nu^{\rm I}_1 (p'-p)^\mu Z'_\mu \left(I^*_{1,n+6} R^*_{1,n+6}-I^*_{1,n+3} R^*_{1,n+3}\right),
\label{interactions}
\eea
}}
where $\Gamma$ and $R~(I)$ are the matrices that diagonalize the CP-even Higgs mass matrix and the CP-even (odd) sneutrino mass matrix, respectively. We assumed that $Y_S$ is diagonal with ${Y_S}_{nn}={Y_S}_{n},$ and $n=1,2,3.$ The  interaction term of $\tilde{\nu}_1^2$ and $hh'$ is not included in the above expression, since it is much smaller than the interactions with $|h|^2$ and $|h'|^2$. However, in our numerical calculations it is taken into account along with other subdominant interactions. Thus, the Feynman diagram of the dominant annihilation channels of $\tilde{\nu}_1$ to CP-even Higgs bosons, SM gauge bosons and three light-neutrinos, $\nu^i_L$, are given in figure \ref{annihilationsnu}. It is clear that the four-point
interaction $\tilde{\nu}_1^2 |h'|^2$ gives the dominant effect for  the annihilation of $\tilde{\nu}_1$.

\begin{figure}[t]
\begin{center}
\includegraphics[width=8cm,height=6cm]{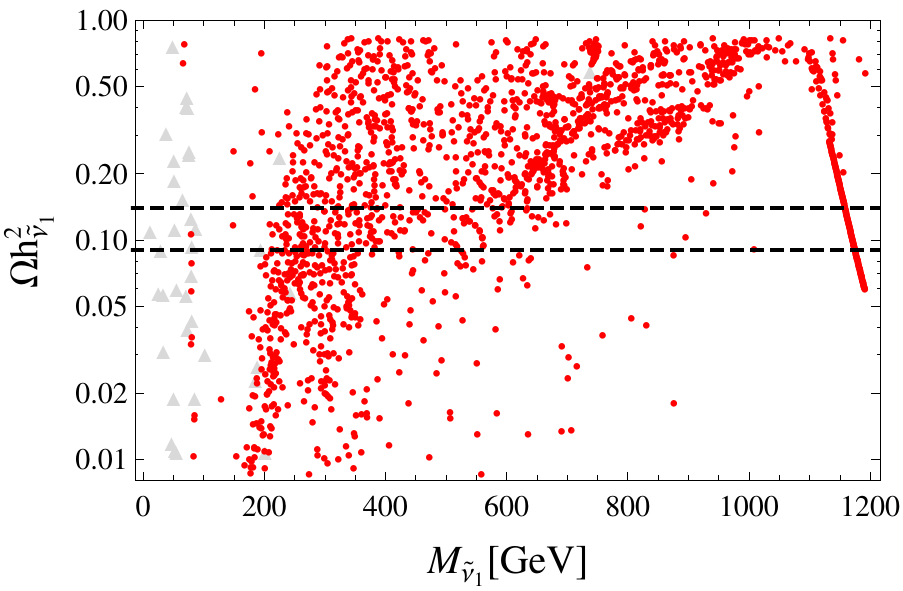}
\caption{The thermal relic abundance of right-handed sneutrino LSP as a function of its mass. The gray triangles denote to the excluded points due to LUX upper bound. Horizontal lines correspond to the Planck limits on DM abundance.} 
\label{MsnuOMGh2snu}
\end{center}
\end{figure}

The relic abundance of the lightest right-handed sneutrino, $\Omega h^2_{\tilde{\nu}_1}$, as function of its mass $M_{\tilde{\nu}_1}$ is presented in figure \ref{MsnuOMGh2snu}, for $m_0,~m_{1/2} \in [100~{\rm GeV},  3~{\rm TeV}]$, $A_0\in[-4,4]$~TeV, $\tan \beta \in[3, 50]$, and $\mu >0$. The observed limits in eq.~(\ref{plancklimit}) and the Higgs mass and gluino mass constraints are imposed. 
As can be seen from this figure, unlike the scenario of lightest neutralino, there are many points with $M_{\tilde{\nu}_1}$ varies from $80$~GeV to $1.2$~TeV are allowed. This wide range of the allowed right-handed sneutrino DM
would rescue the idea of SUSY DM, which faces serious challenges and stringent constraints in the MSSM and also in the BLSSM with neutralino DM candidates.   

\vspace{-0.5cm}
\subsection{Direct Detection Constraints on the Right-handed Sneutrino }
From eq.~(\ref{interactions}),  one can see that the effective scalar interactions of $\tilde{\nu}_1$ with up and down quarks are given by CP-even Higgs bosons ($h$ and $h'$) exchanges. Therefore, the effective coupling $f_q$ in eq.~(\ref{scalar}) is given by 
\be 
f_q \simeq  \frac{g_{\tilde\nu_1 \tilde\nu_1 h}~g_{q\bar{q} h}}{m_{h}^2} + \frac{g_{\tilde\nu_1 \tilde\nu_1 h'}~g_{q\bar{q} h'}}{m_{h'}^2},
\label{fq}
\ee
where these couplings (in the case of the lightest sneutrino being CP-even, $\tilde{\nu}^{\rm R}_1$) are as follows 
\bea
g_{\tilde\nu_1^{\rm R} \tilde\nu_1^{\rm R} h'} &\simeq & \sum_{n=1}^3\frac{g_{_{BL}}^2}{2}  \left(v'_1 \Gamma_{33}\!-\! v'_2 \Gamma_{34}\right)\!-\!\sqrt{2} \mu' \Gamma_{34} R^*_{1,n+6} R^*_{1,n+3} {Y_S}_{n}
\!-\!\Gamma_{33}\! \left(
\sqrt{2} R^*_{1,n+6} R^*_{1,n+3} {T_S}_{n}\!+\!v'_1  {Y_S}^2_{n}
\right),\nonumber\\
g_{\tilde\nu_1^{\rm R} \tilde\nu_1^{\rm R} h}&\simeq& \sum_{n,m,l=1}^3 v_2 \Gamma_{12}\left(R^*_{1,l+3} R^*_{1,m+3} {Y_\nu}_{mn} {Y_\nu}_{ln}-\frac{1}{4} \tilde{g} g_{_{BL}}\right),\\
g_{u\bar{u} h}&=& \frac{m_u}{v} \frac{\Gamma_{12}}{\sin\beta},~~~g_{d\bar{d} h}= \frac{m_d}{v} \frac{\Gamma_{11}}{\cos\beta},\\
g_{u\bar{u} h'} &=& \frac{m_u}{v} \frac{\Gamma_{32}}{\sin\beta},~~~g_{d\bar{d} h'}= \frac{m_d}{v} \frac{\Gamma_{31}}{\cos\beta}.
\eea
From these expressions, one finds that on the one hand the coupling $g_{\tilde\nu_1^{\rm R} \tilde\nu_1^{\rm R} h'}  \sim g_{_{BL}}^2 v' \sim {\cal O}(100)$~GeV and the coupling $g_{q\bar{q}h'}$ is quite suppressed due to very small mixing $\Gamma_{32}/\Gamma_{31}$ and up/down Yukawa coupling, therefore, the effective coupling due to $h'$ exchange in eq.~(\ref{fq}) is of order $Y_d/(m_{h'}^2 \cos\beta) \sim {\cal O}(10^{-7})~{\rm GeV}^{-1}$. On the other hand the coupling $g_{\tilde\nu_1^{\rm R} \tilde\nu_1^{\rm R} h}  \sim Y_\nu^2~ v \sim {\cal O}(10)$~GeV and the coupling $g_{d\bar{d}h} \sim Y_d \Gamma_{11}/\cos\beta \sim 10^{-4}$, therefore, the effective coupling due to $h$ exchange is of order ${\cal O}(10^{-3})~{\rm GeV}^{-1}$. In this case, the effective coupling of $\tilde{\nu}_1$ to proton and neutrino, $f_{p,n}^{\tilde{\nu}_1}$ is about three order of magnitudes larger than the effective coupling of the neutralinos $f_{p,n}^{\tilde{\chi}_1}$. Thus, one would expect a larger spin-independent cross section for sneutrino DM that may even exceed the LUX limits. 

\begin{figure}[t]
\begin{center}
\includegraphics[width=8cm,height=6.2cm]{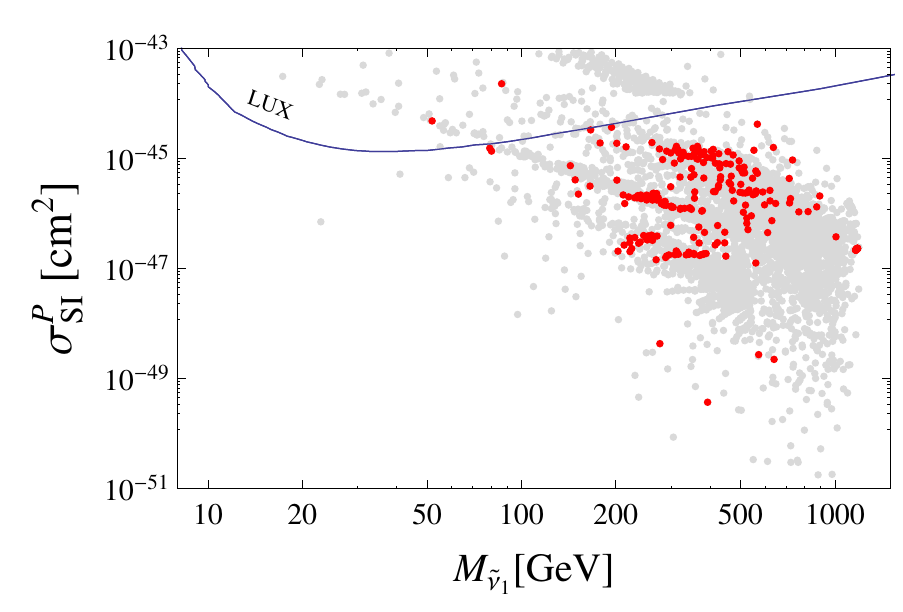}
\caption{The spin-independent cross section of the scattering between the right-handed sneutrino LSP and proton versus its mass. The blue curve is the recent LUX result. The gray points are excluded by the Planck limits on the DM relic abundance.} 
\label{MsnuOMGh2snuLUX}
\end{center}
\end{figure}

In figure \ref{MsnuOMGh2snuLUX}, we show the  spin-independent cross section of the scattering between the $\tilde{\nu}_1$ DM  and proton as a function of its mass. Here we used all the allowed points of the scan in figure \ref{m0mchimsnu} that lead to a viable $\tilde{\nu}_1$ DM. As expected, some points exceed the experimental limits of LUX, however, a significant number of benchmarks predicts cross sections that can be probed in near future. 
\subsection{Right-handed Sneutrino and Galactic Center $\gamma$-rays}
The measured signal from the GC consists of the gamma-ray of photons from annihilating DM particle and a background. Therefore, the differential of total observed gamma-ray flux can be written as
\be\label{Eq:DFSB}
\frac{d\Phi_{\rm tot}}{dE_\gamma}=\frac{d\Phi_{\gamma}}{dE_\gamma}+\frac{d\Phi_{\rm BG}}{dE_\gamma}, 
\ee
where $d\Phi_{\gamma}/dE_\gamma$ is the differential gamma-ray flux generated from the DM,  defined in eq.~(\ref{Eq:DFDM}), and $d\Phi_{\rm BG}/dE_\gamma$ is the Fermi bubbles \cite{Hooper:2013rwa,Fermi-LAT:2014sfa,Dobler:2009xz,Su:2010qj} and the isotropic gamma-ray backgrounds \cite{Abdo:2010nz}.

\begin{figure}[t]
\begin{center}
\includegraphics[width=7.5cm,height=6cm]{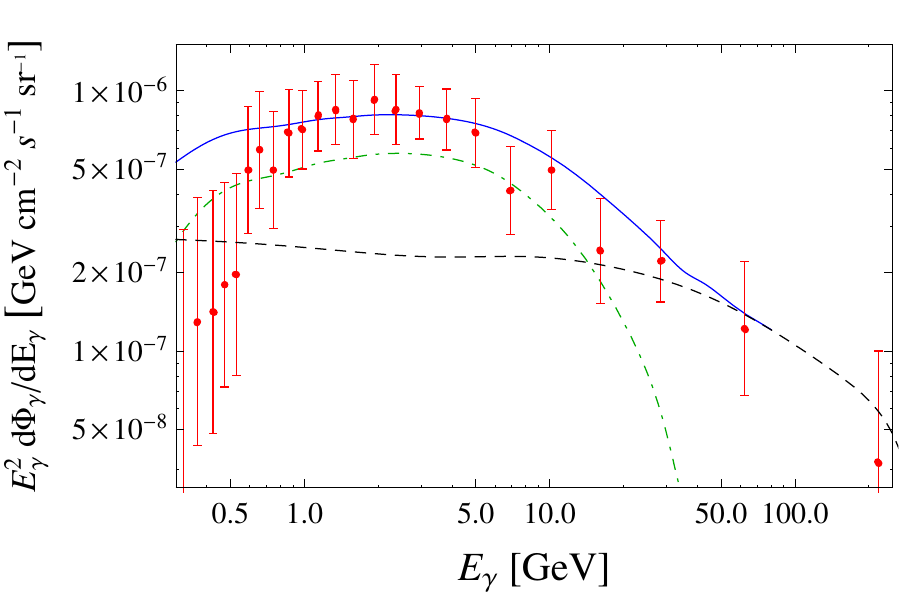}~~~~~\includegraphics[width=7.5cm,height=6cm]{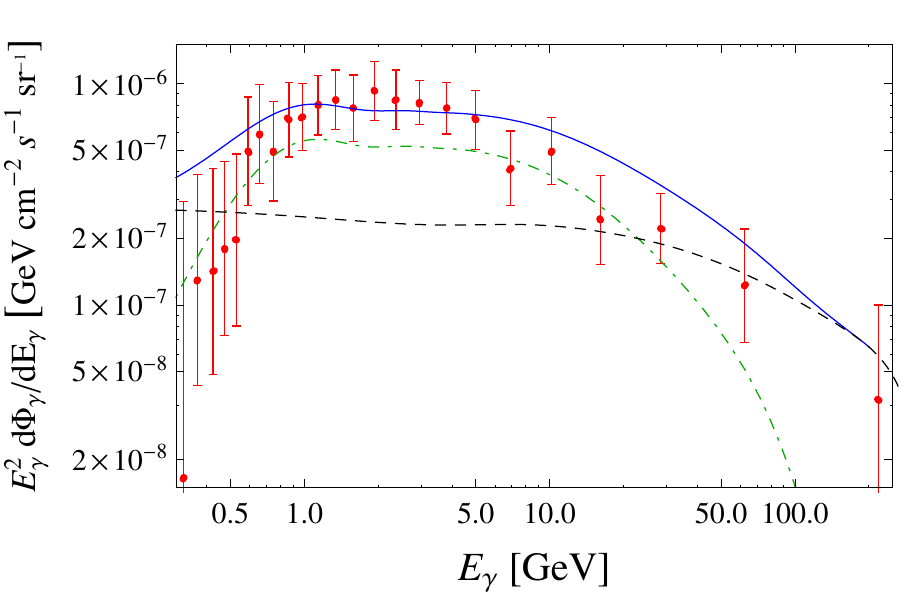}
\caption{The measured spectrum of gamma-rays within the ROI $2^{\circ} \leq |b| \leq 20^{\circ}$ and $|l| \leq 20^{\circ}$ of the GC. The dashed line shows the backgrounds. (Left panel) The gamma-rays spectrum produced for the lightest sneutrino DM annihilation into $WW~(91~\%)$ with $M_{\tilde{\nu}_1} \simeq 80.3$~GeV and total annihilation cross section $\langle \sigma^{\rm ann} v \rangle \simeq 2\times 10^{-26}~ {\rm cm}^3~{\rm s}^{-1}$ (the dot-dashed green curve). (Right panel) The gamma-rays spectrum produced for the lightest sneutrino DM annihilation into $hh~(80~\%)$ and $t\bar{t}~(20~\%)$ with $M_{\tilde{\nu}_1} \simeq 200$~GeV and total annihilation cross section $\langle \sigma^{\rm ann} v \rangle \simeq 5\times 10^{-26}~ {\rm cm}^3~{\rm s}^{-1}$ (the dot-dashed green curve). The solid blue curve shows the sum of the signal and its backgrounds.} 
\label{FERMILAT80GWW680MO}
\end{center}
\end{figure}

In figure \ref{FERMILAT80GWW680MO}, we show the differential flux of gamma-rays originated from the annihilation of the lightest sneutrino DM with masses:  $\sim 80$~GeV (left panel) and $\sim 200$~GeV (right panel), respectively.  These two benchmark points satisfy all astroparticle constraints and the LHC constraints as well.

In case of lightest sneutrino mass $\sim 80$~GeV, the annihilation cross section is dominated by $WW$ channel and given by $\langle \sigma^{\rm ann}_{WW} v\rangle \simeq 2\times 10^{-26}$~cm$^3$~s$^{-1}$. While the annihilation cross section of the lightest sneutrino with mass $200$~GeV is dominated by $hh$ channel $(80~\%)$ and $t\bar{t}$ channel $(20~\%)$ and given by $\langle \sigma^{\rm ann}_{hh} v\rangle \simeq 5\times 10^{-26}$~cm$^3$~s$^{-1}$. 
The Fermi bubbles and the isotropic gamma-ray backgrounds are given by dashed line and our signals are presented by dot-dashed green curve. The sum of signal and background is given by the solid blue curve.\\

As can be seen from this figure,  the observed access by Fermi-LAT at energy $\sim 2 - 5$~GeV can be accommodated by the lightest sneutrino in our two examples. It is worth noting that heavy $\tilde{\nu}_1$ with mass larger than $100$~GeV and dominated annihilation into pair of $WW$ cannot account for these excess, since the peak of the curve is shifted to $E_\gamma \sim 5 - 10$~GeV. 
\section{Conclusions}
In this paper we have analysed the DM sector in the BLSSMIS model, which is a well motivated scenario by neutrinos oscillate and non-vanishing neutrino masses.
The BLSSMIS is an extension of the MSSM obtained by a $U(1)_{B-L}$ extension of the SM gauge group and implementing inverse seesaw mechanism.
We showed that the BLSSMIS offers new cold DM candidates, in addition to the usual MSSM neutralinos. In particular, the extra $B-L$
neutralinos (one $\tilde{B'}$ and two extra higgsinos $\tilde{\eta}_{1,2}$) and the lightest right-handed sneutrino can be the LSP and viable candidates of DM.  
The lightest $B-L$ neutralino is typically a mixed between $\tilde{B'}$ and $\tilde{\eta}_{2}$. It may be a $B'$ino-like if $\mu'$ parameter is much large than the soft parameter $M_{B'}$. 
Due to the large off-diagonal elements in the right-handed sneutrino mass matrix, which are proportional to $\mu'$ and trilinear coupling $A_S \sim$~TeV, one of its eigenvalues is light, 
hence the lightest sneutrino can be of order ${\cal O}(100)$~GeV. 

We have analysed the constraint from the relic abundance observed limits on the three BLSSMIS DM candidates. 
We found that in case of $B'$ino-like DM only three benchmark points are allowed, while a few more points in case of extra $B-L$ higgsino are survived. In case of lightest right-handed sneutrino, it turns out that 
relic abundance constraints can be satisfied for a wide range of parameter space, with DM mass varies from $80$~GeV to $1.2$~TeV. 
We also studied the spin-independent DM scattering cross section with nuclei. We showed that the scattering
cross section of the $B-L$ neutralinos LSP (that satisfies the relic constraints) with a proton is less than the recent LUX bounds, by about two order of magnitudes. However, these bounds exclude some of the right-handed sneutrino allowed points, yet a significant number of benchmark points predicts cross sections that can be probed in near future.
Finally, we pointed out that due to the Majorana type of the $B-L$ neutralino, its annihilation cross section is $p$-wave with no $s$-wave contribution. 
Therefore, its annihilation cross in the galactic halo is of order $10^{-30}~{\rm cm}^3~{\rm s}^{-1}$, which is quite small to account for the GC gamma-ray excess measured by Fermi-LAT.
The right-handed sneutrino is scalar DM, so its annihilation cross section has an $s$-wave contribution. Thus, its value in galactic halo is almost equal to its value at the decoupling limit, $\sim 10^{-26}~{\rm cm}^3~{\rm s}^{-1}$,
thus a lightest right-handed sneutrino with mass ${\cal O}(100)$~GeV can account for the observed GC gamma-ray results, through annihilation to $W W$.

These results indicate that the right-handed sneutrino LSP provides a very compelling example of SUSY DM and it rescues the idea, which faces serious challenges and stringent constraints in the MSSM and also in the BLSSM.

\section*{Acknowledgements}

This work was partially supported by the STDF project 13858, the ICTP grant AC-80 and the European Union's Horizon 2020 research and innovation programme under the Marie Skodowska-Curie grant agreement No 690575



\begin{thebibliography}{99}
\bibitem{Abdallah:2015hza} 
  W.~Abdallah and S.~Khalil,
  Adv.\ High Energy Phys.\  {\bf 2016}, 5687463 (2016)
  [arXiv:1509.07031 [hep-ph]].
\bibitem{Chakraborti:2014fha} 
  M.~Chakraborti, U.~Chattopadhyay, S.~Rao and D.~P.~Roy,
  Phys.\ Rev.\ D {\bf 91}, 035022 (2015)
  [arXiv:1411.4517 [hep-ph]].
\bibitem{Roszkowski:2014wqa} 
  L.~Roszkowski, E.~M.~Sessolo and A.~J.~Williams,
  JHEP {\bf 1408}, 067 (2014)
  [arXiv:1405.4289 [hep-ph]].
\bibitem{Calibbi:2014lga} 
  L.~Calibbi, J.~M.~Lindert, T.~Ota and Y.~Takanishi,
  JHEP {\bf 1411}, 106 (2014)
  [arXiv:1410.5730 [hep-ph]].
\bibitem{Khalil:2006yi} 
  S.~Khalil,
  J.\ Phys.\ G {\bf 35}, 055001 (2008)
  [hep-ph/0611205].
\bibitem{Basso:2008iv}
  L.~Basso, A.~Belyaev, S.~Moretti and C.~H.~Shepherd-Themistocleous,
  Phys.\ Rev.\ D {\bf 80}, 055030 (2009)
  [arXiv:0812.4313 [hep-ph]].
\bibitem{Basso:2009gg}
  L.~Basso, A.~Belyaev, S.~Moretti, G.~M.~Pruna and C.~H.~Shepherd-Themistocleous,
  PoS EPS{\bf{-HEP2009}}, 242 (2009)
  [arXiv:0909.3113 [hep-ph]].
\bibitem{Basso:2010yz}
  L.~Basso, S.~Moretti and G.~M.~Pruna,
  Phys.\ Rev.\ D {\bf 83}, 055014 (2011)
  [arXiv:1011.2612 [hep-ph]].
\bibitem{Basso:2010as}
  L.~Basso, A.~Belyaev, S.~Moretti and G.~M.~Pruna,
  J.\ Phys.\ Conf.\ Ser.\  {\bf 259}, 012062 (2010)
  [arXiv:1009.6095 [hep-ph]].
\bibitem{Majee:2010ar}
  S.~K.~Majee and N.~Sahu,
  Phys.\ Rev.\ D {\bf 82}, 053007 (2010)
  [arXiv:1004.0841 [hep-ph]].
\bibitem{Li:2010rb}
  T.~Li and W.~Chao,
  Nucl.\ Phys.\ B {\bf 843}, 396 (2011)
  [arXiv:1004.0296 [hep-ph]].
\bibitem{Perez:2009mu}
  P.~Fileviez Perez, T.~Han and T.~Li,
  Phys.\ Rev.\ D {\bf 80}, 073015 (2009)
  [arXiv:0907.4186 [hep-ph]].
\bibitem{Emam:2007dy}
  W.~Emam and S.~Khalil,
  Eur.\ Phys.\ J.\ C {\bf 55}, 625 (2007)
  [arXiv:0704.1395 [hep-ph]].
\bibitem{PublicPapers}
  S.~Khalil and S.~Moretti,
J.\ Mod.\ Phys.\  {\bf 4}, 7 (2013)
  [arXiv:1207.1590 [hep-ph]] and
  Front.\ Phys.\  {\bf 1}, 10 (2013)
  [arXiv:1301.0144 [physics.pop-ph]].
\bibitem{Khalil:2010iu} 
  S.~Khalil,
  Phys.\ Rev.\ D {\bf 82}, 077702 (2010)
  [arXiv:1004.0013 [hep-ph]].
\bibitem{BLMSSM-Higgs}
  A.~Elsayed, S.~Khalil and S.~Moretti,
Phys.\ Lett.\ B {\bf 715}, 208 (2012)
  [arXiv:1106.2130 [hep-ph]];
L.~Basso and F.~Staub,
 Phys.\ Rev.\ D {\bf 87}, 015011 (2013)
 [arXiv:1210.7946 [hep-ph]].
\bibitem{Aad:2012tfa} 
  G.~Aad {\it et al.} [ATLAS Collaboration],
  Phys.\ Lett.\ B {\bf 716}, 1 (2012)
  [arXiv:1207.7214 [hep-ex]];
  S.~Chatrchyan {\it et al.} [CMS Collaboration],
  Phys.\ Lett.\ B {\bf 716}, 30 (2012)
  [arXiv:1207.7235 [hep-ex]].
\bibitem{Khalil:2016lgy} 
  S.~Khalil,
  Phys.\ Rev.\ D {\bf 94}, 075003 (2016)
  [arXiv:1606.09292 [hep-ph]].
\bibitem{Guo:2013sna} 
  J.~Guo, Z.~Kang, T.~Li and Y.~Liu,
  JHEP {\bf 1402}, 080 (2014)
  [arXiv:1311.3497 [hep-ph]].
\bibitem{DeRomeri:2012qd} 
  V.~De Romeri and M.~Hirsch,
  JHEP {\bf 1212}, 106 (2012)
  [arXiv:1209.3891 [hep-ph]].
\bibitem{Basso:2012gz} 
  L.~Basso, B.~O'Leary, W.~Porod and F.~Staub,
  JHEP {\bf 1209}, 054 (2012)
  [arXiv:1207.0507 [hep-ph]].
\bibitem{Khalil:2011tb} 
  S.~Khalil, H.~Okada and T.~Toma,
  JHEP {\bf 1107}, 026 (2011)
  [arXiv:1102.4249 [hep-ph]].
\bibitem{Khalil:2008ps} 
  S.~Khalil and H.~Okada,
  Phys.\ Rev.\ D {\bf 79}, 083510 (2009)
  [arXiv:0810.4573 [hep-ph]].
\bibitem{Cerdeno:2004xw} 
  D.~G.~Cerdeno, C.~Hugonie, D.~E.~Lopez-Fogliani, C.~Munoz and A.~M.~Teixeira,
  JHEP {\bf 0412}, 048 (2004)
  [hep-ph/0408102];
  A.~Menon, D.~E.~Morrissey and C.~E.~M.~Wagner,
  Phys.\ Rev.\ D {\bf 70}, 035005 (2004)
  [hep-ph/0404184].
\bibitem{Khalil:2007dr} 
  S.~Khalil and A.~Masiero,
  Phys.\ Lett.\ B {\bf 665}, 374 (2008)
  [arXiv:0710.3525 [hep-ph]];
  P.~Fileviez Perez and S.~Spinner,
  Phys.\ Rev.\ D {\bf 83}, 035004 (2011)
  [arXiv:1005.4930 [hep-ph]].
\bibitem{Haber:1984rc}
  H.~E.~Haber and G.~L.~Kane,
  Phys.\ Rept.\  {\bf 117}, 75 (1985).
\bibitem{Gunion:1984yn}
  J.~F.~Gunion and H.~E.~Haber,
  Nucl.\ Phys.\ B {\bf 272}, 1 (1986)
  [Erratum-ibid.\ B {\bf 402}, 567 (1993)].
\bibitem{ElKheishen:1992yv}
  M.~M.~El Kheishen, A.~A.~Aboshousha and A.~A.~Shafik,
  Phys.\ Rev.\ D {\bf 45}, 4345 (1992).
\bibitem{Guchait:1991ia}
  M.~Guchait,
  Z.\ Phys.\ C {\bf 57}, 157 (1993)
  [Erratum-ibid.\ C {\bf 61}, 178 (1994)].
\bibitem{SUSYRUN1}
  G.~Aad {\it et al.} [ATLAS Collaboration],
  arXiv:1508.06608 [hep-ex];
  A.~Gaz [CMS Collaboration],
  arXiv:1411.1886 [hep-ex];
  I.~Melzer-Pellmann and P.~Pralavorio,
  Eur.\ Phys.\ J.\ C {\bf 74}, 2801 (2014)
  [arXiv:1404.7191 [hep-ex]];
  N.~Craig,
  arXiv:1309.0528 [hep-ph].
\bibitem{gluino search} 
  G.~Aad {\it et al.} [ATLAS Collaboration],
  JHEP {\bf 1504}, 116 (2015)
  [arXiv:1501.03555 [hep-ex]];
  S.~Chatrchyan {\it et al.} [CMS Collaboration],
  Phys.\ Lett.\ B {\bf 725}, 243 (2013)
  [arXiv:1305.2390 [hep-ex]].
\bibitem{Kolb}
E. W. Kolb and M. S. Turner, The Early Universe, Redwood City, USA:
Addison-Wesley (1988) 719 pp., (Frontier in Physics, 70).
\bibitem{Jungman:1995df} 
  G.~Jungman, M.~Kamionkowski and K.~Griest,
  Phys.\ Rept.\  {\bf 267}, 195 (1996)
  [hep-ph/9506380].
\bibitem{Ade:2015xua} 
  P.~A.~R.~Ade {\it et al.} [Planck Collaboration],
  Astron.\ Astrophys.\  {\bf 594}, A13 (2016)
  [arXiv:1502.01589 [astro-ph.CO]].
\bibitem{Belanger:2014vza} 
  G.~Bélanger, F.~Boudjema, A.~Pukhov and A.~Semenov,
  Comput.\ Phys.\ Commun.\  {\bf 192}, 322 (2015)
  [arXiv:1407.6129 [hep-ph]].
\bibitem{Akerib:2015rjg} 
  D.~S.~Akerib {\it et al.} [LUX Collaboration],
  Phys.\ Rev.\ Lett.\  {\bf 116}, 161301 (2016)
  [arXiv:1512.03506 [astro-ph.CO]].
\bibitem{Daylan:2014rsa} 
  T.~Daylan, D.~P.~Finkbeiner, D.~Hooper, T.~Linden, S.~K.~N.~Portillo, N.~L.~Rodd and T.~R.~Slatyer,
  Phys.\ Dark Univ.\  {\bf 12}, 1 (2016)
  [arXiv:1402.6703 [astro-ph.HE]].
\bibitem{Ackermann:2015zua} 
  M.~Ackermann {\it et al.} [Fermi-LAT Collaboration],
  Phys.\ Rev.\ Lett.\  {\bf 115}, 231301 (2015)
  [arXiv:1503.02641 [astro-ph.HE]].
\bibitem{Hooper:2013rwa} 
  D.~Hooper and T.~R.~Slatyer,
  Phys.\ Dark Univ.\  {\bf 2}, 118 (2013)
  [arXiv:1302.6589 [astro-ph.HE]].
\bibitem{Sjostrand:2006za} 
  T.~Sjostrand, S.~Mrenna and P.~Z.~Skands,
  JHEP {\bf 0605}, 026 (2006)
  [hep-ph/0603175].
\bibitem{Navarro:1995iw} 
  J.~F.~Navarro, C.~S.~Frenk and S.~D.~M.~White,
  Astrophys.\ J.\  {\bf 462}, 563 (1996)
  [astro-ph/9508025].
\bibitem{Navarro:1996gj} 
  J.~F.~Navarro, C.~S.~Frenk and S.~D.~M.~White,
  Astrophys.\ J.\  {\bf 490}, 493 (1997)
  [astro-ph/9611107].
\bibitem{Calore:2014xka} 
  F.~Calore, I.~Cholis and C.~Weniger,
  JCAP {\bf 1503}, 038 (2015)
  [arXiv:1409.0042 [astro-ph.CO]].
\bibitem{Fermi-LAT:2014sfa} 
  M.~Ackermann {\it et al.} [Fermi-LAT Collaboration],
  Astrophys.\ J.\  {\bf 793}, 64 (2014)
  [arXiv:1407.7905 [astro-ph.HE]].
\bibitem{Su:2010qj} 
  M.~Su, T.~R.~Slatyer and D.~P.~Finkbeiner,
  Astrophys.\ J.\  {\bf 724}, 1044 (2010)
  [arXiv:1005.5480 [astro-ph.HE]].
\bibitem{Dobler:2009xz} 
  G.~Dobler, D.~P.~Finkbeiner, I.~Cholis, T.~R.~Slatyer and N.~Weiner,
  Astrophys.\ J.\  {\bf 717}, 825 (2010)
  [arXiv:0910.4583 [astro-ph.HE]].
\bibitem{Abdo:2010nz} 
  A.~A.~Abdo {\it et al.} [Fermi-LAT Collaboration],
  Phys.\ Rev.\ Lett.\  {\bf 104}, 101101 (2010)
  [arXiv:1002.3603 [astro-ph.HE]].
\end{thebibliography}
\end{document}